\documentclass[12pt]{article}

\usepackage{graphicx}
\usepackage{subfigure}
\usepackage{authblk}
\usepackage{url}
\newif\ifcomments
\commentstrue
\newcommand{\ignore}[1]{}
\newif\ifrevision
\revisionfalse

\title{On the Role of Conductance, Geography and Topology in Predicting Hashtag Virality}
\author{Siddharth Bora}
\author{Harvineet Singh}
\author{Anirban Sen}
\author{Amitabha Bagchi}
\author{Parag Singla}
\affil{Indian Institute of Technology, Delhi\\
\small
\texttt{\{cs5090253,anirban,mt5100595,bagchi,parags\}@cse.iitd.ac.in}
}

\begin{document}

\maketitle

\begin{abstract}
We focus on three aspects of the early spread of a hashtag in order to
predict whether it will go viral: the network properties of the subset
of users tweeting the hashtag, its geographical properties, and, most
importantly, its conductance-related properties.  One of our
significant contributions is to discover the critical role played by
the conductance based features for the successful prediction of
virality. More specifically, we show that the first derivative of the
conductance gives an early indication of whether the hashtag is going
to go viral or not.  We present a detailed experimental evaluation of
the effect of our various categories of features on the virality
prediction task. When compared to the baselines and the state of
the art techniques proposed in the literature our feature set is able
to achieve significantly better accuracy on a large dataset of 7.7
million users and all their tweets over a period of month, as well as
on existing datasets.
\end{abstract}


\section{Introduction}
\label{sec:intro}

The ability to predict the emergence of virality of a hashtag has
far-reaching consequences in a number of domains. In the commercial
domain knowledge of potentially viral memes may present a variety of
business opportunities but the most important application of a robust
prediction system would be to provide the authorities with the
capability of spotting the emergence of harmful rumors and the
organization of destructive mass action. For example, the role of
Twitter and Blackberry messenger in organizing looting and spreading
rumors was widely observed during the London riots of
2011~\cite{bright-arstechnica:2011}. In 2008 live accounts of the
Mumbai terror attacks went viral on Twitter spreading panic and
providing the attackers with a source of information on police
movements~\cite{oh-rumour:2013}. In both cases early detection of the
spread of a particular kind of meme could have helped arrest it and
prevent disastrous consequences. The nature of such settings is that
they require a prediction system to sift out important (potentially
viral) content from the vast volume of content in the network. The
usual metrics, precision and recall, take on special significance
here. High recall ensures that we do not miss a potentially harmful
meme, while high precision ensures that expensive resources are not
expended on investigating false leads. In this paper we present
prediction algorithms for Twitter, although we feel that techniques
using similar features can be used on other social networks (like
those derived from cellular messaging or calling data) as well and so
will have wide applicability.

Studying the factors that lead to virality in online systems has been
an important theme in the literature over the last few years,
pioneered by Leskovec, Backstr\"om and Kleinberg's study of the
evolution of memes in blogs~\cite{Leskovec_2009}. The phenomenon of
virality wherein a particular meme--a theme or topic or piece of
content--spreads widely through the network has, in particular,
attracted a lot of attention. What has been largely missing is
prediction that focuses on the efficacy of structural properties of
meme diffusion. The only efforts of this nature are the two attempts
by Weng et. al.~\cite{weng&al13,weng&al13a} and their work, on a user
network of about 0.6 million users is not conducted on a
satisfactorily large scale.  We undertake a classification task on a
large dataset that we compiled containing 7.7 million Twitter users
and all their tweets over a 35 day period. For every hashtag in our
set that appears in a certain number of tweets (we call this number
the {\em prediction threshold} and set it to 1500 in this paper) we
try to predict, at the point when it reaches this number, if it will
reach a number of tweets about one order of magnitude larger. This
latter number, we call it the {\em virality threshold}, is set to
10,000. In what follows we will refer to this task as {\em virality
  prediction}. The features that we use for this classification task
deliberately ignore the content of the tweeted hashtags, focusing
instead on three aspects: network topology, geography, and the
isoperimetric properties of meme growth as expressed through the
conductance of the set of users tweeting on a hashtag. This should
{\em not} be seen as a claim that these characteristics are more
important than semantic aspects of the topics under study. We omit
using semantic features to maintain a focus on the predictive power of
structural properties of hashtags. It is not our intention to present
a ``best possible'' prediction algorithm; such an algorithm would
undoubtedly include semantic features along with our structural and
geographic features. Our intention is primarily to demonstrate that there
is significant information contained in our new features based on
conductance, geography and network characteristics of early adopters,
and that this information can be effectively used in the important
task of predicting virality.

We view the evolution of the hashtag's spread across the network as a
graph process and derive a number of network-based features. The
geographical spread of hashtags is also used for virality prediction
for the first time, enabled by a methodology that is successfully able
to tag 90\% of active users with their time zones. We define features
based on the isoperimetric quantity conductance that is known to have
a strong relationship to mixing times in random
walks~\cite{guruswami-unpub:2000}, and find that these features
greatly help improve our predictions. We use the criteria of
information gain to identify the top features from each category
providing us greater insight into the effectiveness of various
features for characterizing virality.

We perform an extensive experimental evaluation of our proposed set of
features 
for the task of predicting virality.  Our
experimental results clearly demonstrate the effectiveness of our
features for this task as well as their supremacy over existing
approaches proposed in the literature. 
We also present some
preliminary experimental analyses of virality prediction in individual
geographies which corroborate our findings in the larger dataset.

\paragraph{Organization} We survey the literature in
Section~\ref{sec:related}, following that with a description of our
dataset and how it was compiled in Section~\ref{sec:dataset}. Our
task definition can be found in Section~\ref{sec:task}, followed by a
detailed discussion of our features in Section~\ref{sec:features} and
the results of our algorithms on two datasets in
Section~\ref{sec:expt}. Finally we conclude with a discussion of the
significance of conductance and some future directions in
Section~\ref{sec:conclusions}.

\section{Related Work}
\label{sec:related}
The problem of predicting which memes will grow in popularity has
attracted some amount of attention recently. To save space we ignore
the research on detecting popularity after it has been established
and focus on the research on identifying memes that will go viral {\em
  before} they have already spread far and wide. Several aspects of
this problem have been studied. Wu and Huberman, in an early paper,
highlighted the importance of novelty: new memes override older
ones~\cite{Fang}. In a similar vein, Weng et. al.~\cite{Weng2} argued that 
finite user attention coupled with the structure of the network can help
identify the ultimate popularity of a meme. Emotional,
textual and visual features have been studied as drivers of
virality~\cite{berger,guerini,guerini-socialcom:2013,totti-websci:2014,Hansen}. In
the context of viral marketing, Aral and Walker showed that
personalization of promotional messages helps make particular products
more ``contagious''~\cite{Aral}.

While all these aspects are undoubtedly important, our work falls into
a different category of research which views the proliferation of
memes as a kind of contagion process on a network and relies on
spatial and temporal properties of the early evolution of this process
to identify potentially viral memes. Leskovec et. al.'s work on memes
falls into this category, positing a temporal growth model for viral
memes~\cite{Leskovec_2009}. With the growth of microblogging it was
natural that such phenomena be investigated on Twitter and Kwak
et. al.~\cite{Kwak_2010} performed the first analysis of information
spreading on Twitter at scale. Subsequently, several groups of
researchers have investigated the structural properties of rapidly
spreading themes by looking at the spread as a cascade or a tree
(e.g. Ghosh and Lerman~\cite{ghosh-wsdm:2011}). Others tried to find
the extent of external (or exogenous) influence on information
diffusion (e.g. Myers et. al.~\cite{myers-kdd:2012}). Szabo and
Huberman tried to predict the long-term popularity of a meme based on
its nascent time series information~\cite{Szabo}. Kitsak
et. al. argued that the most efficient spreaders are those that exist
within the core of the network and the distance between such spreaders
often governs the maximum spread of topics~\cite{Kitsak}. The factors
that govern a tweets ability to draw retweets, a possible precursor to
virality, has been studied by Suh et. al.~\cite{Suh}.  On the modeling
front, Romero et. al. focused on the local dynamics of hashtag
diffusion~\cite{Romero_2011}. Lermann et. al.~\cite{Lermann} propose an
approach to predict popularity of news items in Digg using
stochastic models of user behavior. Rajyalakshmi et. al. defined a
stochastic model for local dynamics with implicit competition that was
found to generate the global dynamics of a Twitter-like
network~\cite{rajyalakshmi2012topic}.  Finally we mention the work
that is the major take off point for our current paper: Ardon
et. al. conducted a study of information diffusion on a large data set
and established that conductance and geographical spread could
significantly discriminate between topics that went viral and those
that did not~\cite{Ardon}. Drawing on ideas from this work we create
suite of features and show that they can be used effectively to
predict virality.

Efforts to use machine learning techniques to predict virality have
recently begun to appear in the literature. Jenders et. al. use
features very different from ours, relying mainly on sentiment-related
features, to predict which tweets will go viral via the process of
retweeting on a small data set of 15,000
users~\cite{jenders-msnd:2013}. Zaman et. al. use a Bayesian approach
to predict which tweets will generate large retweet trees on a small
set of 52 tweets~\cite{zaman-aas:2014}. Ma et. al. combine a set of
textual features with some network-based features to predict the
number of users tweeting a hashtag in subsequent time intervals, a
task somewhat different from ours since we focus on predicting an
eventual ascent to a threshold-based
virality~\cite{ma-jasist:2013}. In a major recent work, Cheng
et. al. tried to predict which photo resharing cascades will grow past
the median cascade size using a variety of features that included
demographic, structural and temporal
information~\cite{cheng-www:2014}. \ifrevision {\bf Rev id 1
  $\rightarrow$} \fi The primary difference between that work and ours
is that we do not work with a cascade model, but look at the spread of
hashtags as a diffusion, i.e., the use of hashtag by a particular user
need not be {\em explicitly} attributable to the prior use of that
hashtag by another user in the neighborhood. This makes our work
incomparable with that of Cheng et. al.~\cite{cheng-www:2014} and
gives it a different flavour.  \ifrevision {\bf $\leftarrow$ End Rev
  id 1} \fi Closer to our approach in conception if the work of Weng
et. al. who showed in a sequence of two papers that inter-community
spread in the early life of a Twitter meme can be used to predict
which meme will go viral and which will
not~\cite{weng&al13,weng&al13a}. The main work against which our
results should be compared is~\cite{weng&al13a} where a
straightforward attempt to predict virality is made, as opposed
to~\cite{weng&al13} where a related but slightly different multi-label
classification task is defined. Our current work overcomes some of the
severe shortcomings of~\cite{weng&al13,weng&al13a}. Firstly, we
perform prediction on a user set of size 7.7 million and view their
interconnections as a directed graph. In~\cite{weng&al13,weng&al13a}
the user set has size only 0.6 million and consists of edges only
between those users who follow each other, i.e. only bidirectional
edges. Secondly, the small size of their network allows Weng
et. al. to run community detection algorithms which are prohibitively
expensive to run. We show that on our much larger data set, to the best of our
knowledge the largest on which virality prediction tasks have been run
so far, by leaving out community-based features and using
computationally tractable features based on geography and conductance
we are able to give better quality predictions than those
in~\cite{weng&al13a}. We also show that our feature set performs
better than the community-based features of Weng
et. al.~\cite{weng&al13a} on their own data set, even though we do not
have the geography information for their users.

\section{Dataset and Methodology}
\label{sec:dataset}

\subsection{Dataset description}
\label{sec:dataset:description}
\begin{figure}[htbp]
 \centering
   \includegraphics[scale=0.20]{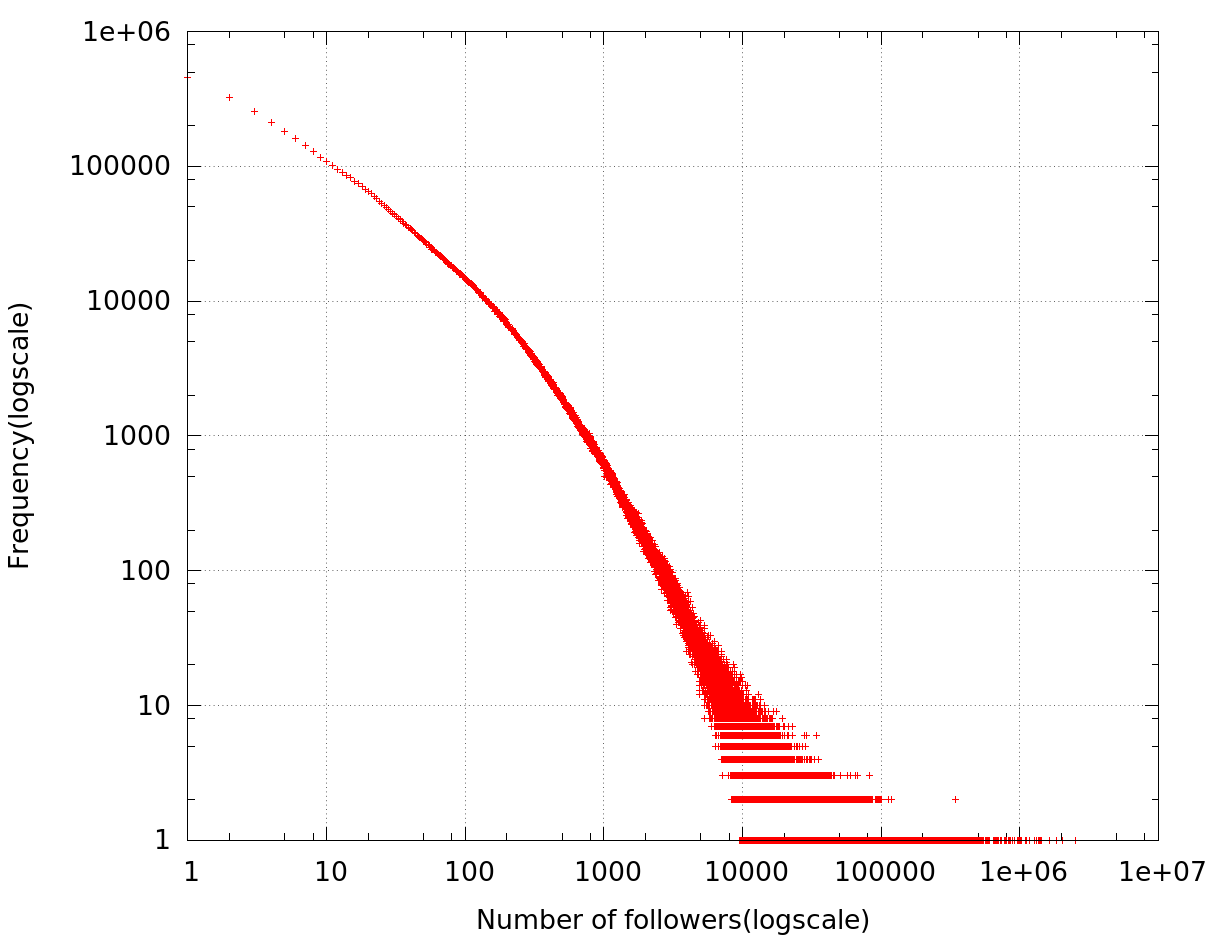}
 \caption[Different Test]{Plot of the distribution of the Number of followers}
 \label{fig:followers}
\end{figure}

Our dataset is a complete snapshot of all tweets posted by 7.7 million
users of Twitter between 27th March 2014 and 29th April 2014. We also
have the follower-following information of all these users and have
built the data set (as explained below) to ensure that these users
form a strongly connected subnetwork of Twitter, i.e., for each pair of
users $u, v$ there is a directed path $u$ to $v$ and a directed path
from $v$ to $u$. Rather than filtering out tweets based on topical memes,
we crawled {\em all} the tweets posted by our user set in the time
window with a view to capture all the phenomena, viral and
non-viral, present in the network at the time.
\begin{table}[htbp]
\small
\centering
\begin{tabular}{||l|r||}
\hline
\hline
Users & 7,695,882\\
\hline
Average no. of followers & 450 \\
\hline
Users who tweet at least once & 3,008,496 \\
\hline
Filtered hashtags & 8,793,155\\
\hline
Tweets for filtered hashtags & 220,012,557 \\
\hline
Hashtags with $\geq$ 10,000 tweets & 177 \\
\hline
\hline
\end{tabular}
\caption{Dataset statistics}
\label{tbl:dataset}
\end{table}
Table~\ref{tbl:dataset} contains some basic statistics of our data
set. 
The follower distribution curve of our data (see
Figure~\ref{fig:followers}) follows the now
familiar power law with a drooping tail that has been widely reported
in the literature. 

From the tweets of our user set we extracted
all the hashtags used, filtering some persistent generic hashtags
out (as explained in Sec~\ref{sec:dataset:processing}.)

Additionally we geotagged all the users who have tweeted at least once
with their location information to an accuracy of 98\%, not down to
the city level but down to the time zone level using the time zones
that Twitter requires users to fill through a dropdown menu at the
time of account creation. There were 141 time zones found in our data
set. We note that Twitter provides annotated time zones that are more
numerous than the 40 different time offsets from UTC that are
generally used for timekeeping. For example, although all of India
follows a single time (Indian Standard Time: UTC +5:30), Twitter gives
Indian users four choices--Chennai, New Delhi, Kolkata,
Mumbai--corresponding to four major metropolitan centres, all of which
are marked as GMT +05:30 in the dropdown.

\subsection{Retrieving the data from Twitter}
\label{sec:dataset:crawl}

A seed set of approximately 108K users was shared with us by Pranay Agarwal
who devised a methodology
for differentiating users whose tweets were informative from those
users who used Twitter as a forum for
chatting~\cite{agarwal-iitd:2013}. The seed set comprised users whose
tweets were generally informative in nature, ensuring that our user
set is focussed on a subset of users who transact emergent memes.
We extracted the
follower and following information of this seed set and computed the
strongly connected component that turned out to be of size 64K. We
then queried Twitter for all the users who were either following or
followed by these 64K users. This gave us an initial set of 9,188,701
users.

We then used the ``GET followers'' method of the Twitter API to
extract the follower and following information of these 9.1 million
users. We were successful in extracting this information for 8,379,871
users. The remaining users had their accounts suspended or their
follower and following information was protected. We computed the
strongly connected component of this reduced set. Its size was found
to be 8,047,811. 

Using the ``GET user\_timeline'' method from Twitter's REST API (v
1.1) we collected tweets from these 8.04 million users' timelines.
The GET user method provides the 3200 most recent tweets of each user
and so in order to build a dataset of at least a month's duration we
repeated the crawl after 10 days. The first crawl began on 18th April
and ended on 20th April. The second crawl began on 30th April and
ended on 2nd May. The two crawled sets were then combined to finally obtain all
the tweets posted from 27th March to 29th April 2014, without
duplication or omission. This process of combination involved
discarding 26,624 of the users whose high frequency of tweeting made
it impossible to guarantee that we had captured all their tweets in
this 35 day time span. 
After removing these 26,624
users we recomputed the strongly connected component of the remaining
network and found it to have size 7,695,882. These 7,695,882 users
with their interconnections formed our final user network.

\subsection{Filtering the hashtags}
\label{sec:dataset:processing}

On examination we found that the most frequently occurring hashtags in
the tweet dataset created were generic hashtags that
persist on twitter with a very high frequency of occurrence e.g. \#rt, \#tlot,
\#win, \#giveaway and \#jobs. In order to filter out these hashtags and obtain a set of
hashtags that are fresh in our time span and thereby refer to emergent
phenomena we filtered out all the hashtags
that had more than 5 tweets in the first 12 hours of our time
duration. 
\begin{table}[htbp]
\small
\centering
\begin{tabular}{||l|r||}
\hline
\hline
Hashtag & Tweets \\ \hline \hline
 bundyranch & 395,179\\ \hline
 nashto2mil & 152,657\\ \hline
  votekatniss & 140,031 \\ \hline
  votetris & 135,206\\ \hline
  epnvsinternet & 104,575\\ \hline
\hline
\hline
\end{tabular}
\caption{Top 5 hashtags in our dataset}
\label{tbl:hashtags}
\end{table}
Table~\ref{tbl:hashtags} shows the 5 most popular hashtags in our
dataset after filtering along with the number of tweets for each of
them. The first one, ``bundyranch'', refers to a major US news story of
the time. The second one commemorates a Twitter celebrity Nash Grier's
follower count reaching 2 Million. The third and fourth are related to
voting prior to the MTV Movie awards that were held on 13th April
2014, and the fifth one ``epnvsinternet'' is part of a mass action by
civil organizations in Mexico to oppose a proposed legislation. These
five demonstrate anecdotally that our filtering process does generally
capture emergent topics on Twitter  and also
that endogenous topics generated from within Twitter like
``nashto2mil'' and exogenous topics like ``bundyranch'' and
``votekatniss'' are both present in our dataset.

\subsection{Geolocating the users}
\label{sec:dataset:geography}

Twitter users have the option of specifying their time zone, their
country and their location. The first two of these are populated from
a dropdown menu and the third is entered as a string and hence is
often hard to map to an actual geographical location. These three
fields are part of the user's profile and are embedded in the JSON
object containing the user's tweet, which is where we extracted them
from. This JSON object also contains geo-coordinates from tweets
posted from GPS-enabled devices whose users have allowed this
information to be shared but we found that to be a rare occurrence and
not of much use.
\begin{table}[htbp]
\small
\centering
\begin{tabular}{||l|r||}
\hline
\hline
Time Zone & Users \\ \hline \hline
Eastern Time (US \& Canada)    &  967,849\\ \hline
Central Time (US \& Canada) &     636,541\\ \hline
Pacific Time (US \& Canada) &     549,611\\ \hline
London & 395,738\\ \hline
Quito& 212,584\\ \hline
\hline
\hline
\end{tabular}
\caption{Top 5 time zones in our dataset}
\label{tbl:timezones}
\end{table}
On examination we found that timezone was a widely present attribute,
missing only from 630,696, i.e., 11.9\% of the 5.3 million users who had
posted at least one tweet in our initial set of 9.1 million
users. About a third of these users, 217,443, had tweeted from
GPS-enabled devices and so we were able to map them to time zones
using the Google Time Zone API. For the remaining users we extracted
their location string and tried to map it to a time zone by using common
substring heuristics to match their location strings with popular
cities and with location strings of users that have their time zones
set. This helped us locate another 118,488 users.  Finally, we were
left with 294,765 users. To these users we assigned the time zone in
which the maximum number of their neighbors were located. Testing this
heuristic on users whose location was known we obtained a 48\%
success rate. So, in summary, we were able to correctly geolocate all
but 294,765 users out of 5.3 million, i.e., 94.4\% of our users. The
remaining 5.6\% were tagged with a heuristic that we expect to perform
correctly about half the time. Even if we consider only the subset of
3 million users who tweet once in our 35 day period and assume that
the 294,765 users whose location we guess all lie in this subset, we
see that 90\% of our users are correctly tagged and the remaining 10\%
are tagged by a heuristic that has 48\% accuracy. We note that there
have been several research efforts made to geolocate users but
they have been either at the country level
e.g.~\cite{kulshrestha-icwsm:2012} or at a fine-grained level of tens
of kilometers e.g.~\cite{mcgee-cikm:2013}. We adopt a simpler strategy
here to achieve geo-location at the intermediate granularity of
Twitter time zones since geo-location is not the primary focus of our
paper. As reported we find that the accuracy we achieve using our
simple, and computationally efficient, methods is significant and good
enough for our purpose. Table~\ref{tbl:timezones} lists the top five
time zones by population in our user set.
We see that along with what we
traditionally understand by time zones like ``Eastern Time (US \&
Canada)'', we also have individual cities like ``London'' and
``Quito.''

\section{Prediction Task}
\label{sec:task}

The goal of our study is to examine how successfully we can
discriminate viral topics from non-viral ones based on their early
spreading pattern.  In this section we make this goal concrete. We
undertake here a classification task whose objective is {\em to
  predict at a particular, early, point in the spread of a hashtag
  whether that specific hashtag will, in the future, enter the class
  that we define as viral.}
\paragraph{Defining virality}  Various
definitions of virality have been used in the literature. One of them
is based on calling a topic viral if it is among the top $k$\% (for
some small value of $k$) of all the topics ranked by their total
number of tweets. This definition has been used by Weng et
al.~\cite{weng&al13a} for values of $k$ starting at 10. A drawback of
this definition is that, because of its relative nature, it is
non-monotonic. In other words, a topic may become viral at a given
point of time, but then be declared non-viral at a later point in its
life when some other topics have surpassed it in terms of number of
tweets and it is no longer in the top $k\%$.  While it is definitely
true that a viral topic ceases to be viral after some time, to make
this percentile-based definition stick we would also have to provide
some kind of time window within which the topic must remain in the top
$k$\%.  In order to avoid such complications we decided to use an
alternate definition based on an absolute threshold, i.e., we say a
topic has become viral if its total number of tweets cross a certain
given threshold $M$. We call this the {\em virality threshold}. We
used $M=10,000$ in our experiments. This number has no intrinsic
significance. It is based on the sizes of the spreads of various
hashtags in our data set. Note that while we saw in
Table~\ref{tbl:hashtags} that the 5 largest spreading tags had more
than 100,000 tweets, choosing a virality threshold of 10,000 gives us
only 6.29\% of hashtags. This top 6.29th percentile that we consider
viral is significantly smaller than the 10th percentile that is taken
as viral by Weng et al.~\cite{weng&al13a}.

\paragraph{Prediction threshold} Since we are
interested in predicting the future spread based on early history, we
decided to extract features from the first $n=1500$ tweets. We call
this value $n$ the {\em prediction threshold}. In other words, for
each hashtag, we examined its spread in the network upto the point the
$1500$th tweet containing that hashtag was posted and extracted
various features based on this early spread. All the topics which did
not cross the $n=1500$ tweet mark were ignored for the purpose of our
study. The total number of hashtags that we were left with was 2810,
of which 177 crossed the virality threshold.

\begin{figure}
\centering
\includegraphics[width=0.50\textwidth]{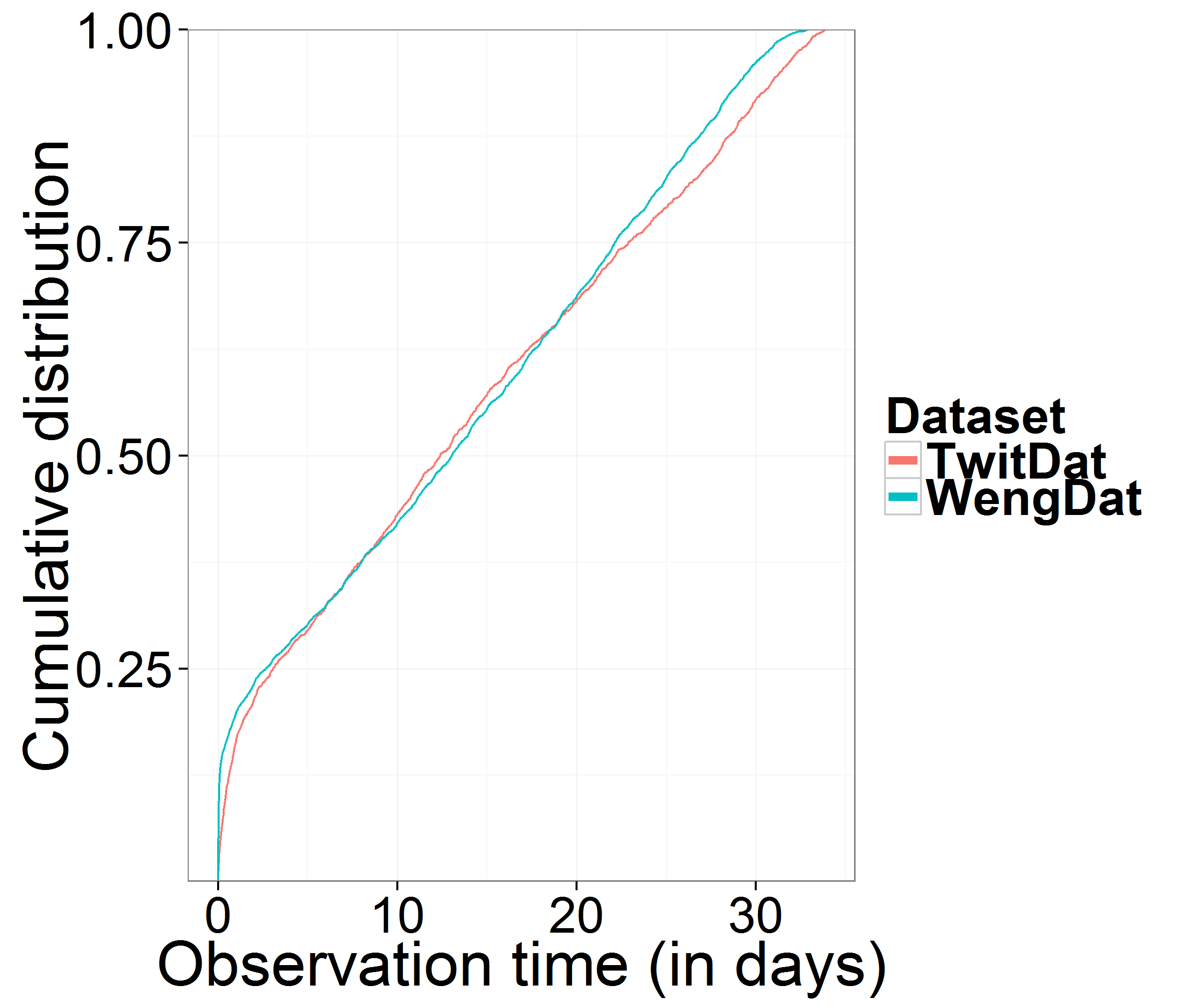}
\caption{Time taken to 
reach the prediction threshold.}
\label{fig:time-dist-predict-thresh}
\end{figure}

Choosing a very small value of $n$ may not give us sufficient
information about the topic spread but a very large value of $n$ will
make the prediction meaningless as the peak would already be reached
(or be very close). So, it is important to choose $n$ in an
appropriate manner. Further note that the choice of appropriate $n$
would also depend on the size of the dataset. For a large dataset $n$
would be higher compared to a smaller dataset simply because of the
sheer volume of tweets generated in the network. To put in context our
choice of $n=1500$, another similar recent study by Weng et
al.~\cite{weng&al13a} has used $n=50$ tweets as their threshold. The
size of their dataset (in terms of the number of users) is about 5\%
of the size of the dataset that we use (595K nodes vs 7.7 million
nodes).
In order to show that these two prediction threshold values, Weng
et. al's 50 and our 1500 are similar we plotted the time taken to
reach $n=1500$ tweets for a particular hashtag in our dataset and the
time taken to reach $n = 50$ tweets in the dataset of Weng
et. al.~\cite{weng&al13a}. 
In
Figure~\ref{fig:time-dist-predict-thresh} we have time on the $x$-axis
and the fraction of hashtags that take at most a given time to reach
the prediction threshold for our dataset ({\it TwitDat}) theirs ({\it WengDat}).
The overlapping nature of the two curves clearly demonstrates the similarity in 
the behavior of the topics at the two prediction thresholds for the two
datasets, respectively. 
The median time taken to reach the prediction
threshold is $12.38$ days for our data set and $13.03$ days for Weng
et al's. We interpret these results to mean that the amount of
information available at our prediction threshold for our dataset
closely aligns with the amount of information available in the other
prediction threshold for the other major study, and hence our results
can be compared with theirs.

\section{Our features}\label{sec:features}
The main contribution of this paper is the definition of a set of
novel features that are critical to the task of predicting
virality. We propose a number of new features and argue that one of
the most important aspects of early hashtag growth is the rate of
change of the conductance of the subset of nodes that are tweeting the
hashtag. Our prediction algorithm is also the first to incorporate a
set of geography based features. Apart from our conductance and
geography based features we also use a set of temporal features (we
call these ``evolution based features'') and a set of features that
capture the network characteristics of the early tweeters of a
hashtag. In total we have experimented with $29$ different features. 
The features are listed in Table~\ref{table:feature-description}.

We will refer to the users
who tweet on a hashtag up to the prediction threshold as {\em
  adopters} of that hashtag. Of these a special category are what we
call {\em self-initiated adopters} who tweet on a hashtag before any
of the users that they follow do so. In some places we will use the
term ``topic'' interchangeably with ``hashtag''. The term
``geography'' will be used to denote Twitter time zones as described
in Section~\ref{sec:dataset} e.g., if we say ``the average
number of users in a geography is X'' we mean that the average number
of users in a Twitter time zone is X. Further, we will use the term
{\em weakly connected component} in the way it has come to be
understood i.e. given a directed graph if we treat each directed edge
as undirected and compute the connected components of the transformed
graph, then each of these connected components is known as a {\em
  weakly connected component} of the original directed graph.

\begin{table*}[!htbp]
\centering
\scriptsize{
\begin{tabular}{|p{0.32\linewidth}|p{0.6\linewidth}|}
\hline
{\bf Name} & {\bf Description} \\
\hline
\multicolumn{2}{|c|}{\bf Evolution based features} \\
\hline
{\em NumOfAdopters}    &  Number of adopters who tweeted on the hashtag  \\
{\em NumOfRT}  &   Number of retweets (RT) on tweets within the prediction threshold \\ 
{\em NumOfMention}  &   Number of user mentions (@) in tweets within the prediction threshold \\ 
{\em TimeTakenToPredThr}  &  Growth rate of the hashtag measured in terms of time taken to reach prediction threshold   \\
\hline
\multicolumn{2}{|c|}{\bf Network based features} \\
\hline
{\em HeavyUsers}  &  Number of adopters with at least $3000$ followers   \\ 
{\em NumFolAdopters}  &   Total number of followers of adopters  \\ 
{\em NumOfEdges} & Number of edges in the network spread, i.e., the subgraph induced by the set of adopters  \\ 
{\em Density} & Subgraph density     \\ 
{\em SelfInitAdopters} & Number of Self-initiated adopters \\ 
{\em SelfInitAdoptersFollowers} & Total follower count of Self-initiated adopters \\ 
{\em RatioOfSingletons} & Ratio of Self-initiated adopters to number of adopters \\ 
{\em RatioOfConnectedComponents} &   Ratio of number of weakly connected components to number of adopters  \\ 
{\em LargestSize} & Size of the largest weakly connected component \\ 
{\em RatioSecondToFirst} & Ratio of sizes of the second largest to the largest weakly connected components \\
\hline
\multicolumn{2}{|c|}{\bf Geography based features} \\
\hline
{\em InfectedGeo}   &    Number of infected geographies   \\ 
{\em RatioSelfInitComm}   &    Fraction of Self-initiated geographies   \\ 
{\em RatioCrossGeoEdges}   &    Fraction of edges across geographies in the induced subgraph of adopters  \\ 
{\em AdoptEntropy}   &    Adoption Entropy measures the distribution of adopters across geographies and is defined as $-\sum\limits_{i} a_i\log a_i$, where $a_i$ is the fraction of adopters in each geography $i$  \\ 
{\em TweetingEntropy}   &    Tweeting Entropy measures the distribution of tweets across geographies and is defined as $-\sum\limits_{i} t_i\log t_i$, where $t_i$ is the fraction of tweets in each geography $i$  \\ 
{\em IntraGeoRT}   &    Fraction of retweets occurring between users from the same geography   \\ 
{\em IntraGeoMention}   & Fraction of user mentions occurring between users from the same geography   \\
\hline
\multicolumn{2}{|c|}{\bf Conductance based features} \\
\hline
{\em CummConductance}  & Conductance of the subgraph induced by the set of adopters   \\
{\em Conduct'\_k, $k=~\{20,50,100,250\}$}  &  First derivative of conductance for different values of smoothing parameter $k$  \\
{\em Conduct''}  &  Second derivative of conductance  \\
{\em Conduct'\_stdev, Conduct''\_stdev}  &  Standard deviation of first and second derivative of conductance \\
\hline
\end{tabular}
}
\caption{Feature descriptions. All features have been captured with respect to tweets within the prediction threshold}
\label{table:feature-description}
\end{table*}
\subsection{Feature Categories}

We divided our features in the following four categories: 1) {\em
  Evolution based} features capture very basic analytics of the
hashtag's evolution such as number of adopters, number of retweets,
number of user mentions and growth rate of the topic. Since these are
very simple features we will be using them to generate baselines. 2)
{\em Network based} features include various network characteristics
of the adopters of the topic in terms of their followers, density,
self-initiated adopters and weakly connected component based
features. 3) {\em Geography based} features capture the geographical
properties of the spread such as number of infected geographies, intra
and inter geography features, number of self-initiated adopters in
each geography etc. 4) {\it Conductance based} features, though based
on network properties, have been put in a separate category due to
their prime importance for the task of characterizing virality. These
include conductance as well as its first and second derivative. Next
we discuss the features in each of the above categories.

\begin{figure*}[htbp]
\centering
\begin{subfigure}[Evolution based features (E)]{
\includegraphics[width=0.45\textwidth]{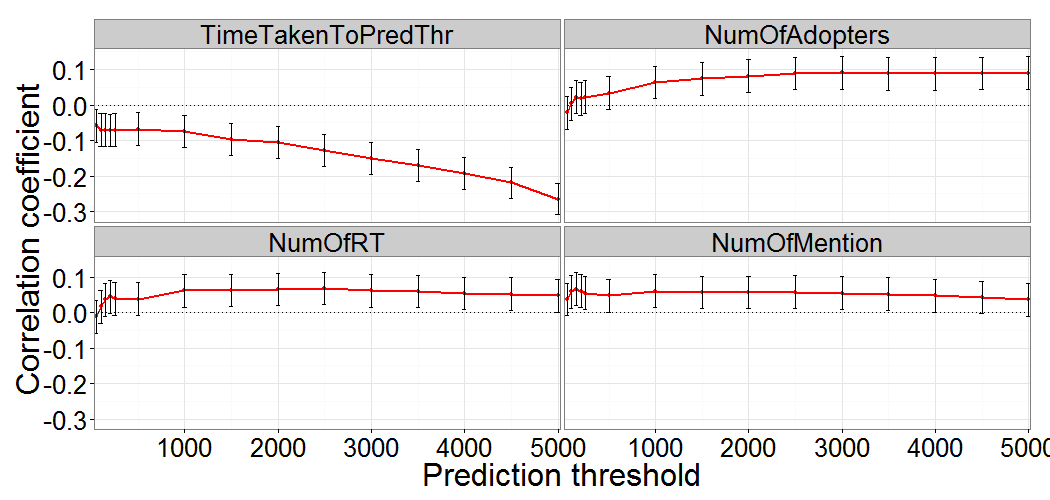}
\label{fig:top_feature_E}
}
\end{subfigure}
\begin{subfigure}[Network based features (N)]{
\includegraphics[width=0.45\textwidth]{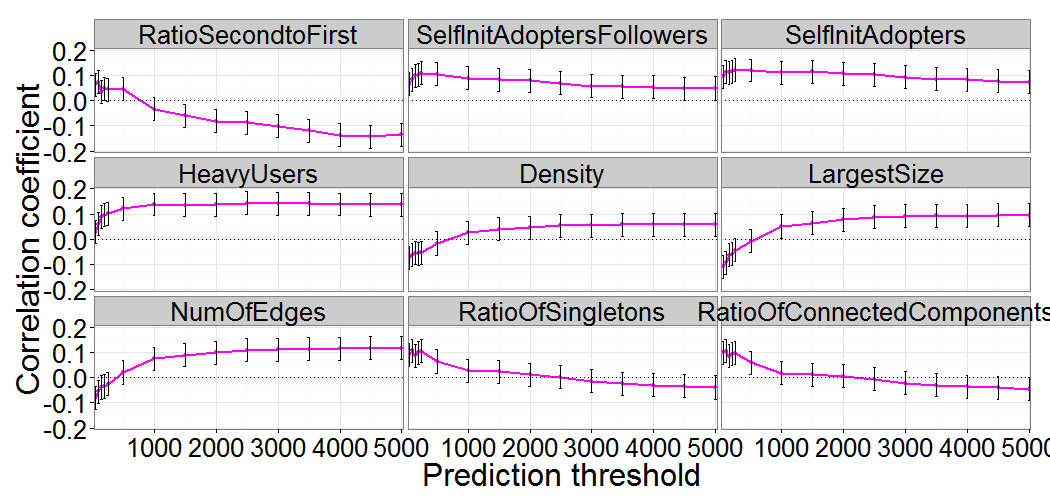}
\label{fig:top_feature_N}
}
\end{subfigure}
\begin{subfigure}[Geography based features (G)]{
\includegraphics[width=0.45\textwidth]{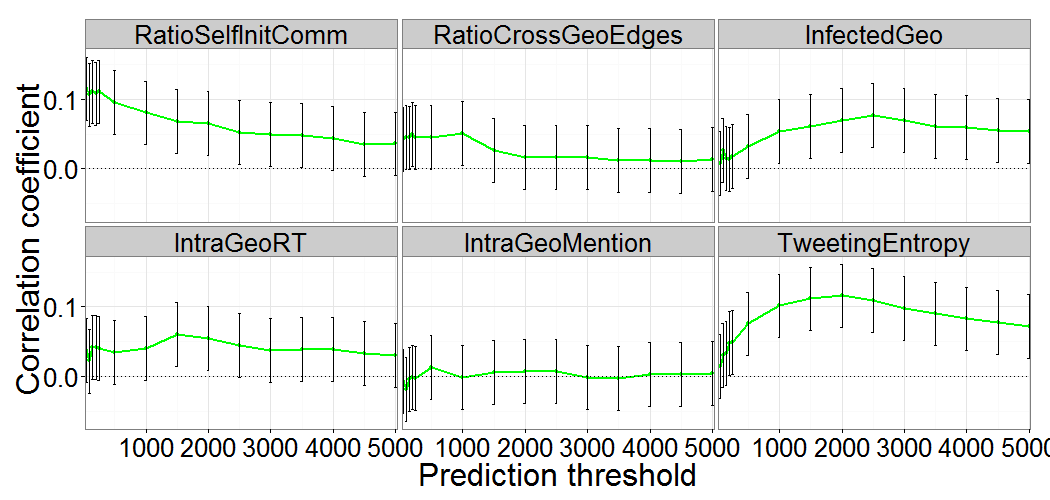}
\label{fig:top_feature_G}
}
\end{subfigure}
\begin{subfigure}[Conductance based features (C)]{
\includegraphics[width=0.45\textwidth]{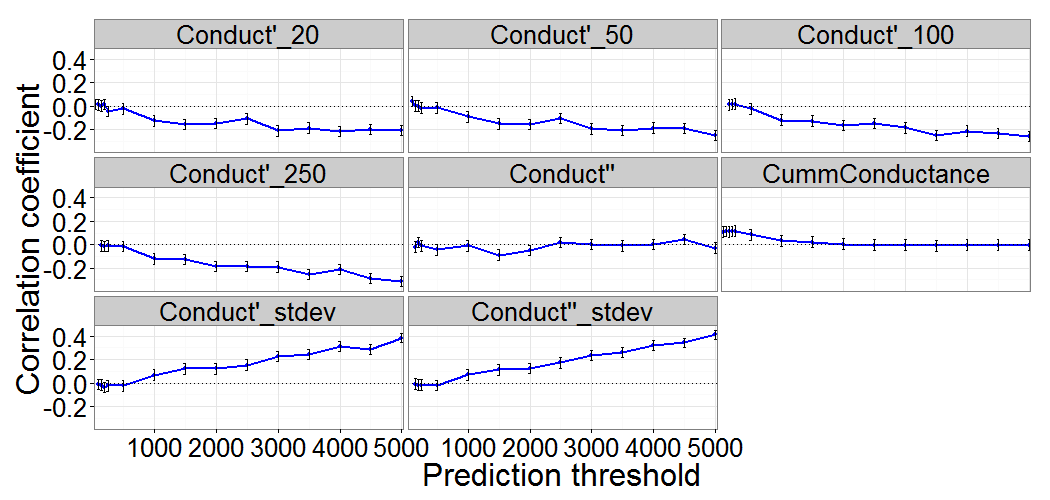}
\label{fig:top_feature_C}
}
\end{subfigure}
\caption{Change in correlation coefficient with increase in number of tweets. Bars indicate the $95\%$ confidence interval}
\label{fig:corr-coeff}
\end{figure*}

\subsubsection{Evolution-based features}
\label{sec:features:categories:evolution}
These features include basic characteristics about the topic evolution
and include the following $4$ features: 1) {\em Number of Adopters}
2) {\em Number of Retweets} 
3) {\em Number of User Mentions} 
4) {\em Growth Rate} defined as the time taken to reach the 
prediction threshold. We group these in two sets, denoting the first
and the fourth features as the set E1, and the second and third as E2. We note that the
first two features
have been used before by Weng
et. al.~\cite{weng&al13,weng&al13a}. Growth Rate was used, along with
a number of variations thereof, by Cheng
et. al.~\cite{cheng-www:2014}. Zaman et. al.~\cite{zaman-aas:2014}
used the number of retweets and other aspects of retweeting as
features for their prediction task while Jenders
et. al.~\cite{jenders-msnd:2013} used number of user mentions as a
feature. For us, as mentioned earlier, this set of features will be
used to create non-trivial baselines. The first baseline will use only
the features E1 while the second baseline will use all four
features. The details of how these baselines will be used are
discussed in Section~\ref{sec:expt}.

We plotted the change in {\em Spearman's rank} correlation coefficient, 
measured between feature values and total hashtag growth, with increasing
number of tweets for all hashtags that had at least 5000 tweets (see
Figure~\ref{fig:top_feature_E}). For the feature set E we find the number
of adopters and number of RTs are positively correlated with hashtag
growth but the correlation levels out, while the time to prediction
threshold is highly negatively correlated and continues to grow in
this direction. This latter observation is similar to that of Cheng
et. al.~\cite{cheng-www:2014} who observed that successful cascades
get many views in a short amount of time.

\subsubsection{Network-based features}
We used the following $10$ network based features (divided into 3
sub-categories) for our study.  The first subset includes features
based on adopters and their connections: 1) {\em Number of Adopters
  with Heavy Following} where a user with at least 3000 followers is
said to have a heavy following (recall that the average number of
followers in our dataset is 450). This discriminative significance of
this feature has been discussed by Ardon et. al.~\cite{Ardon}. A
related but somewhat different feature, the average authority of
users, was used by Ma et. al.~\cite{ma-jasist:2013}. 2) {\em Number of
  Followers of Adopters}. 3) {\em Number of Edges in the Network
  Spread} 4) {\em Subgraph Density} defined as the ratio of the number
of edges to the number of nodes in the network spread. Versions of
these three feature have been used by Cheng
et. al.~\cite{cheng-www:2014} and Jenders
et. al.~\cite{jenders-msnd:2013} and also discussed by Ardon
et. al.~\cite{Ardon}.

The second subset includes $3$ features based on self-initiated
adopters, i.e., adopters with no neighbors who have adopted the same
hashtag before the prediction threshold. These include 1) {\em Number
  of Self-Initiated Adopters} 2) {\em Follower Count of Self-Initiated
  Adopters} 3) {\em Ratio of Self-Initiated Adopters to Number of
  Adopters}. We note that the cascade setting of Cheng
et. al.~\cite{cheng-www:2014} involves, by definition, just one
``root'' whereas we can have any number of self-initiated adopters,
reflecting the critical difference between that setting and the
setting we study here.

Lastly, we used $3$ weakly connected component based features. These
include 1) {\em Ratio of Number of Weakly Connected Components to
  Number of Adopters} 2) {\em Size of the largest Weakly Connected
  Component} 3) {\em Ratio of the Sizes of the Two Largest Weakly
  Connected Components}. Ardon et. al.~\cite{Ardon} have posited a
merging phenomenon in the growth of a hashtag to virality: initially
the growth of the meme takes place in small separate clusters that
begin merging for those memes that are moving towards virality, but
remain separate for those memes that are not. These three features
attempt to quantify this process. To the best of our knowledge they
have never been used for this kind of prediction task before.

For the network features of feature set N, we find that the
correlation coefficient is generally positive and significant and
remains stably so (Fig.~\ref{fig:top_feature_N}). An important
deviation is the ratio of the second largest to the largest component
which is negatively correlated with hashtag growth. This negative
correlation continues to increase as the number of tweets seen
increases reflecting the fact that components merge to find a giant
component as virality approaches.

\subsubsection{Geography-based features}

Taking user geographies (time zones as defined by Twitter that we
extracted (see Section~\ref{sec:dataset})) into account, we
were able to define a set of 7 features. 1) {\em Number of Infected
  Geographies} i.e. geographies with at least one tweet about the
topic.  2) {\em Fraction of Self-Initiated Geographies} i.e. fraction of
geographies where the first user tweeting was self-initiated.  3) {\em
  Fraction of Edges across Geographies} i.e., fraction of total edges
whose end points lie in different geographies. We used 1) {\em
  Adoption Entropy} and 2) {\em Tweeting Entropy} across geographies
as two of our features. We captured intra-geography activity using the
following two features: 1) {\em Fraction of Intra-Geography Retweets}
and 2) {\em Fraction of Intra-Geography Mentions}. The fractions refer
to the fraction of total number of retweets and mentions to the
intra-geography retweets and mentions, respectively. Of these
features, we note that the fraction of edges across geographies has
been highlighted as a discriminative metric by Ardon
et. al.~\cite{Ardon}. The rest are similar in flavor to the
community-based features used by Weng
et. al.~\cite{weng&al13,weng&al13a}, except we use time zone as our
notion of community here. The exception to this is the feature {\em
  Fraction of Self-Initiated Geographies} which is used for the first time
here. We note that our current work is the first time, to the best of
our knowledge, that geographical information is being used for
virality or meme growth prediction.

Looking at the correlation coefficient evolution of the feature set G
we find that the number of infected geographies is positively
correlated with successful topics and remains stably so
(Fig.~\ref{fig:top_feature_G}). Notable here is that the tweeting entropy
displays a high correlation with hashtag growth.

\subsubsection{Conductance-based features}
Given a graph $G=(V,E)$ and a subset of nodes $S \subseteq V$, the
conductance $\phi(S)$ of the set $S$ is defined as the ratio of the
number of edges outgoing from $S$ (i.e. follower links of the nodes in
the set $S$) that land outside $S$ i.e.:
\begin{equation}
\phi(S)=\frac{|\{(u \rightarrow v): u \in S, v \in V\backslash S\}|}{|\{(u \rightarrow v): u \in S, v \in V\}|}
\end{equation}
\ifrevision {\bf Rev id 2 $\rightarrow$} \fi
Conductance is a isoperimetric quantity that has been shown to be
closely related to mixing times of random walks in
graphs~\cite{jerrum-siamjc:1989}. In a diffusion setting more general
than a random walk, Chierichetti et. al. showed that the time taken
for a rumor to spread through a network can be characterized in terms
of the conductance of the
graph~\cite{chierichetti-stoc:2010}. Empirical evidence linking
conductance with diffusion in graphs was provided by Ardon
et. al.~\cite{Ardon} who found that the conductance of viral memes
undergoes a sharp dip as they approach virality. To visualize this in
the network setting, we can think of it this way: When a topic goes
viral it saturates the structural community enveloping it, hitting the
low conductance boundary of that community. With this in mind we chose
to investigate a set of conductance based features for our prediction
problem.  
\ifrevision {\bf $\leftarrow$ End Rev id 2} \fi

We used 3 types of conductance based features in our study: the
conductance, and its first and second derivatives. To calculate the
first derivative at prediction threshold $n$, the following
methodology was used. Let us define a {\em time instant} as the
occurrence of a tweet event. Let $c_i$ denote the conductance value at
the $i^{th}$ time instant. Then, for a given smoothing parameter $k$,
the conductance derivative (w.r.t. number of time instants elapsed) is
defined as
\[\frac{\Delta c}{\Delta t}=\frac{(c_{n} - c_{n-k})}{(t_{n} - t_{n-k})}\]
In words, the conductance derivative is the ratio of difference in the
conductance value at the prediction threshold ($n$) and the conductance value
$k$ tweets prior to the prediction threshold to the difference in their
corresponding timestamps. The second derivative of conductance is calculated in
a similar manner using the values of the first derivative. For the first derivative, 
we used the value of $k$ as $20,50,100$ and $250$. For the second derivative, we used
the first derivative values at $k=50$ and $k=100$. We also
measure the standard deviation of the first and second derivatives of
the conductance over the last 100 tweets before the prediction
threshold is reached.
This resulted in a total of $8$ conductance based features
(conductance, $5$ first derivative based features and $2$ second
derivative based feature).

The importance of the change in conductance as a feature becomes clear
when we look at the way the correlation coefficients of the
derivatives of conductance evolve with the number of tweets
(Fig.~\ref{fig:top_feature_C}). In particular the standard deviation
of both the first and second derivatives grow continuously, reaching a
very high value of 0.4 when 5000 tweets have been seen. This strong
correlation foreshadows the striking effect on prediction results that
this feature set is found to have (see Section~\ref{sec:expt}).

\paragraph{The intuition behind Conductance based features} Our
conductance-based features should be compared to the ``first surface''
feature of Weng et. al.~\cite{weng&al13,weng&al13a}, or the ``border users'' of Ma et. al.~\cite{ma-jasist:2013}, which are simply
the number of ``uninfected'' neighbors of users who have tweeted a
hashtag i.e. neighbors who have not yet tweeted that hashtag. This
``first surface'' (and the similarly defined ``second surface'') is
subtler than the features in the flavor of ``number of neighbors''
used by Jenders et. al.~\cite{jenders-msnd:2013} and Cheng
et. al.~\cite{cheng-www:2014} which do not distinguish between
neighbors that have already propagated the meme and those that have
not. But conductance goes a step further. Conductance has been widely
used to {\em measure} the quality of communities produced by community
detection algorithms (see e.g.~\cite{leskovec-www:2010}
or,~\cite{harenberg-wics:2014}) and even been shown to be tightly
related to the clustering coefficient of a
graph~\cite{gleich-kdd:2012}. In view of this, the correct way of
interpreting the conductance based features is by viewing the
diffusion process in its early stages as moving inside communities. At
the point of virality, a hashtag saturates each of these communities,
i.e., it reaches the boundary of the community and the conductance
falls since the edges leaving a community are a small fraction of the
total edges of the community, the majority of the edges being pointed
inward. Hence the ratio of the outward edges to the total edges is a
much more important feature than simply the number of outward edges
because it captures how clustered a certain set of nodes is. This
intuition is borne out by the fact that the derivatives of conductance
are highly correlated with topic growth (as we saw in
Fig.~\ref{fig:top_feature_C} and by the strong impact these features
have on the quality of prediction (see Sec.~\ref{sec:expt}).

\paragraph{Information gain of our features}
We further quantified the efficacy of our proposed features for the
purpose of prediction by computing the {\em information gain} of each
feature. This metric, roughly speaking, reflects the amount of
information the knowledge of a particular feature value of the
evolution of a hashtag upto the prediction threshold gives us about
the class--viral or non-viral--to which the hashtag belongs. Given two
random variables $X$ and $Y$, the {\em information gain of $Y$ with
  respect to $X$} captures the reduction in entropy of $Y$ given
$X$. Information gain is a symmetric metric that is also known as the
{\em mutual information} between $X$ and $Y$ and is represented as
$I(X,Y)=H(Y) - H(Y|X)$. Here $H(Y|X)$ is the conditional entropy of
$Y$ given $X$ and is defined analogous to the entropy, now using the
conditional distribution of $Y|X$.  Recall that entropy is defined as
$H(Y) = -\sum_{i} P(Y=y_i) \log(P(Y=y_i))$ where $P(Y=y_i)$ is the
probability that $Y$ takes the $i^{th}$ state/value.  Intuitively,
information gain captures the amount of information that knowing $X$
can give us about $Y$. 

Table~\ref{table:info-gain} shows the top $2$ features from each
category based on their information gain. Conductance-based features
were found to have the highest information gain among all the features
across all the categories. The top feature in this category is the
first derivative of the conductance which implies that the
speed of the spreading process is a key indicator of its
eventual success. This validates our earlier thesis about conductance
and its properties being very important features for characterizing
virality. 
The highest value of information gain that we have is
$0.04527$ for the first derivative of the conductance. This is a
non-trivial value but is relatively small, which provides us a
quantitative measure of the hardness of the prediction problem.  We
also note that computing the information gain of individual features
does not reveal the entire story since it does not take their
dependence into account. The joint effect of features comes out in our
experimental results (Section~\ref{sec:expt}.)

\begin{table*}[th!]
\footnotesize
\begin{center}
\begin{tabular}{|l|c|c||l|c|c|}
\hline
{\bf Feature} & {\bf Set} & {\bf Info.} & {\bf Feature} & {\bf Set} & {\bf Info.} \\
\quad  &   &   {\bf Gain}     & \quad & & {\bf Gain} \\  
\hline
1. Growth Rate & E & 0.02424     &  1. Tweeting Entropy & G & 0.01249 \\ 
2. No. of Adopters & E & 0.00979       &  2. No. Of Infected & G & 0.00938 \\
\quad  &   &       & \quad Geographies & & \\  
\hline
1. No. of Adopters & N & 0.01175  & 1. 1st Derivative & C & 0.04527 \\ 
\quad with Heavy Following  &   &       & \quad of Conductance (k=50)  & & \\            
2. Number of Edges & N & 0.0099      & 2. Stdev of 2nd Derivative & C & 0.03526 \\ 
\quad  &   &       & \quad of Conductance & & \\  
\hline
\end{tabular}
\caption{Top Ranking Features Based on Information Gain}
\label{table:info-gain}
\end{center}
\end{table*}

\section{Experiments}\label{sec:expt}
The goal of our experiments was to answer the following questions: a)
How effective is each set of features (and the combinations thereof) defined
in Section~\ref{sec:features} in predicting hashtag virality? b) What is 
the impact of changing prediction threshold on virality prediction? c) How 
does our approach compare with existing approaches on existing datasets?
In order to answer these questions, we used the features defined in 
Section~\ref{sec:features} in a machine learning classification and learned 
a model to predict which hashtags go viral. Specifically, the task was 
to predict whether the number of tweets containing a hashtag will cross the 
virality threshold or not given its feature values at the prediction threshold. 

To answer the first two questions, we experimented on the dataset compiled
by us as detailed in Section~\ref{sec:dataset}. In order to answer the last
question, we compared our approach with that of Weng et al.~\cite{weng&al13a}
on their dataset, under the same experimental conditions. 

\subsection{Experimental Setup}
We will refer to our dataset detailed in Section~\ref{sec:dataset} 
as {\it TwitDat}. For experiments on this data, we used the methodology 
defined in Section~\ref{sec:task} for defining which hashtags are declared 
to be viral. The prediction threshold was chosen to be $n=1500$.  Only those 
hashtags which crossed the $n=1500$ mark were used for training and testing 
purposes. This left us with a total of $2810$ hashtags. Only about $6.3\%$
of these hashtags were found to cross the virality threshold of 10,000.

\paragraph{Algorithms}
We refer to our feature based approach for predicting virality as CGNP
(Conductance Geography and Network topology based Predictor). We
experimented with using various combinations of our feature
sub-categories defined in Section~\ref{sec:features} i.e. 1) Evolution
Based (E) which was used to generate two baselines 2) Network Based
(N) 3) Geography Based (G) 4) Conductance Based (C). When using CGNP
with a certain subset of feature categories, we will append the names
of categories used as features. For example, CGNP(E+N) means that we
are using evolution based and network based features only. We compare
CGNP using various feature combinations with the following $3$
baselines:

{\bf Random:} This is the na\"{\i}ve algorithm which randomly (with 0.5 probability)
predicts a hashtag to be viral.

{\bf CGNP(E1):} This is the feature based prediction using the very
basic evolution features, i.e., number of adopters and number of
retweets.  We will refer to these set of features as E1. This is used
as a baseline because of the very intuitive nature of these features for
prediction and their prior use for prediction in the past literature
(see Sec.~\ref{sec:features:categories:evolution} for details).

{\bf CGNP(E):} This is the feature based prediction using all the four
evolution based features described in
Section~\ref{sec:features}. These features enhance the feature set E1
but are still basic and have been used before (see
Sec~\ref{sec:features:categories:evolution}) and so can be thought of
as a baseline.

\paragraph{Learning Methodology}
We compare various prediction algorithms across two primary metrics:
{\em AUC}, i.e. area under the Precision Recall curve and {\em
F-measure}.  We will also report {\em Precision} and {\em Recall}.  
For all our experiments, we used Random Forests with $500$ trees as our 
learning algorithm. For training of each decision tree, 
$\lceil \log_2 {\#features} \rceil$ number of random features are used, where ${\#features}$ 
is the total number of features considered in the learning algorithm. 
We performed 10 fold cross validation over a random 
split of the data for training and testing purposes. 

We briefly describe the evaluation metrics used: AUC, Precision, Recall 
and F-measure. 
The class probabilities assigned to each test data example by the learning algorithm are subsequently compared with a threshold $\theta$, to transform the probability values to binary outputs (1, if the class probability is greater than $\theta$ and 0, otherwise).
These predicted labels for the examples are compared
with the corresponding actual class labels to get the number of 
true positives ({\em tp}), false positives ({\em fp}), true negatives ({\em tn}) and
false negatives ({\em fn}), where positive refers to the virality class. Then,
Precision=$\frac{tp}{tp+fp}$, Recall=$\frac{tp}{tp+fn}$, and 
F-measure, or {\em F1-score} is the harmonic mean of Precision and Recall, i.e.,
F-measure=$\frac{2.Precision.Recall}{Precision+Recall}$.
The Precision-Recall curve is obtained by varying the value of the threshold, $\theta$.
AUC is calculated as the numerical approximation of area under this curve. Thus, AUC
gives a threshold-independent measure of classifier performance and is often used in 
cases of datasets with high class imbalance~\cite{davis2006relationship}.

Note that class distribution is very skewed for our dataset with class size 
ratio for the virality and non-virality classes being close to 1:15. Using 
the ideas from literature to deal with high class
imbalance~\cite{drummond&holte03}, we undersampled the majority 
(non-viral) class at a rate of about 0.3 to bring the class size ratio to
1:4.5. Undersampling was done only on the training folds and test distribution 
was kept as is. 
\subsection{Effect of Feature Sets}
To answer the first question, Table~\ref{table:results} shows the
values of AUC, F-Measure, Precision and Recall for the baselines used
as well as various combinations of feature categories for CGNP. The
best performing feaure combination has been highlighted in bold for
each of the metrics. Random has the highest recall of $50\%$ but has
an extremely low precision. CGNP(E) is the strongest baseline
algorithm among the 3 compared.  There is a gradual improvement in
both AUC and F-measure as more sophisticated features are
added. Adding both network (E+N) and geography (E+G) based features
leads to some improvement in prediction results, effect of geography
being somewhat more than that of network based features. Combining
them together (E+N+G) does not lead to any further improvement in
results which probably means that the two feature categories are
capturing similar effects.

There is a significant improvement in Precision, Recall, F-measure
as well as AUC over the baseline using conductance based features. 
Conductance results in both F-measure and AUC going up by more than $7\%$ 
over the baseline. This points to a very strong efficacy of conductance 
features in predicting virality. This observation is in line with the 
correlation graphs and information gain numbers presented in Section~\ref{sec:features}. 
Adding network based and geography based features leads to a further
improvement in results of about $1.5\%$ for F-measure (E+N+C) and up to $3.5\%$ 
for AUC (E+N+G+C). This means that though conductance is the most
effective feature for prediction, there is some additional signal
captured by network and geography based features for this task. The
best performing feature combination for F-measure is (E+N+C) and
for AUC is (E+N+G+C).

Both our F-measure and AUC numbers appear somewhat on the lower
side. This is because virality prediction is an extremely difficult
task for prediction. Nevertheless, what we really care about is
capturing early on a reasonable fraction of hashtags which would go
viral with some accetable number of false positives (hashtages
predicted viral which were actually not). With the (E+N+G+C) model, we
have a recall of close to $40\%$ with a precision of about
$33\%$. This means that we are able to capture $2$ out of every $5$
hashtags that go viral, while paying the cost of sieving through $3$
hashtags for every truly viral hashtag output by the system.  This
seems reasonable considering the difficutly of the task and a highly
skewed positive class ratio of less than $7$ in $100$.
\begin{table}[ht!]
\centering
\small{
\begin{tabular}{|l|c|c|c|c|}
\hline
{\bf Algorithm} & {\bf Precision} & {\bf Recall} & {\bf F-meas.} & {\bf AUC} \\
\hline
Random    &    6.30  &  {\bf 50.0}  & 11.19  & 6.30 \\
CGNP(E1)  &    13.51  &  35.03  & 19.49  & 14.9 \\ 
CGNP(E)   &    30.00 &	25.42 &	27.52 &	18.5 \\ \hline
CGNP(E+N) &    21.69 &	38.98 &	27.88 &	20.7 \\
CGNP(E+G) &    29.12 &	29.94 &	29.53 &	20.9 \\
CGNP(E+C) &    {\bf36.65}	& 33.33 &	34.91 &	26.2 \\
CGNP(E+N+G) &  22.65 &	36.72 &	28.02 &	20.3 \\
CGNP(E+G+C) &  30.08 &	45.19 &	36.12 &	28.0\\
CGNP(E+N+C) &  31.4 &	42.94 &	{\bf36.28} &	28.2 \\
CGNP(E+N+G+C) & 32.7 &	38.98 &	35.57 &	{\bf 30.0}\\
\hline
\end{tabular}
}
\caption{Results comparing CGNP using various feature combinations and baselines on TwitDat (all values in \%)}
\label{table:results}
\end{table}
\subsection{Effect of Prediction Threshold}
\begin{figure}[htbp] \centering
\includegraphics[scale=0.45]{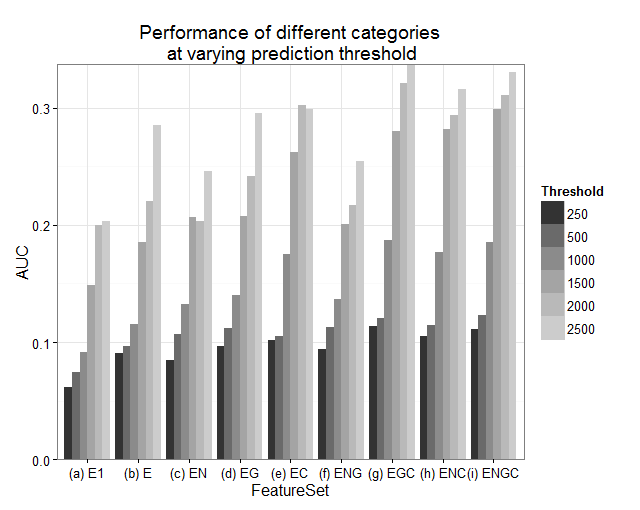}
\caption{AUC versus Prediction Threshold}
\label{fig:varying-predict-thresh}
\end{figure}

To answer the second question, we analyzed the performance of our algorithms
with varying prediction threshold. As explained in Section~\ref{sec:task}, a small
value of the prediction threshold may not give good prediction results,
whereas a large value of the threshold may not be very useful since we are not 
able to make the prediction early enough in the hashtag evolution history. 
Figure~\ref{fig:varying-predict-thresh} plots the variation in AUC for different 
feature sets of CGNP. At any given value of prediction threshold, adding more 
sophisticated features helps improve the performance further (barring few 
minor exceptions). The most improvement is obtained by adding conductance based 
features as observed earlier.

As expected, the performance of all the models improves with
increasing prediction threshold. The maximum rate of increase is seen
when prediction threshold goes from 1000 to 1500 (which is the value
of prediction threshold in rest of our experiments) after which the
gains seem to taper off. This justifies our choice of prediction
threshold by showing that 1500 tweets is {\em the earliest point} at
which a certain quality of prediction can be achieved.

\subsection{Comparison on Existing Datasets}
To answer the third question, we experimented on the dataset of 595K
users compiled by Weng et al.~\cite{weng&al13a}, made available on
their
website.\footnote{\url{http://carl.cs.indiana.edu/data/\#virality2013}} We
will refer to their dataset as {\it WengDat}. 
For our results to be directly comparable with Weng et al.'s approach
on their dataset, we used {\em their} definition of virality, i.e., a
hashtag is declared to be viral if it lies in the top $10\%$ of all
the hashtags in a ranking based on the total number of tweets for each
hashtag at the end of the observation period. We used their
prediction threshold as used by them, i.e., $n=50$, and 
their learning setting as theirs i.e., Random Forests with 500 trees
and 10 fold cross validation over a random split of the data. We did
not perform any undersampling on this dataset.  We compared our
feature combinations with Weng et al.'s feature based approach.  We
refer to their approach as WFVP (Weng Feature based Virality
Predictor).  For WFVP, we directly use the results reported by them in
their paper using their best combination of features.

Table~\ref{table:results-wengdat} compares the performances of WFVP and
CGNP using various feature combinations. Weng et al. already demonstrated 
the superior performance of their approach over a number of non-trivial 
baselines so we report their final results here. We do not have AUC numbers 
for WFVP as they are not reported by Weng et al. We did not have geography 
information for their dataset so we used CGNP with only the three feature
groups E, N and C.
First, note that as on TwitDat, adding sophisticated features to CGNP
helps improve performance. The most improvement is obtained by adding
conductance based features as was the case with TwitDat.  Compared to
WFVP, we are able to achieve a significantly high recall at the cost
of a smaller loss in precision. In a scenario where it is
important not to miss a potentially viral topic (as is the case with
many of our motivating applications), obtaining a high recall becomes
important. Note that overall, our precision-recall combination results
in an F-measure which is 4.5\% more than the best results reported by
Weng et al. A more fine grained comparison with WFVP throws additional
light. Using only evolution and network based features, CGNP performs
worse than WFVP in terms of F-measure.  We attribute this primarily to
the community based features used by Weng et al. which have been shown
to be quite effective for prediction. Significantly, once conductance
based features are added, CGNP starts outperforming WFVP, even when we
do not include network based features. Effectively, conductance is
able to make-up for the lack of community based features for the task
of virality prediction (and in fact, performs better).
Further, we note that our conductance based features are {\em
  local} in the sense that they can be computed by examining the
relevant portion of the network where the hashtag is currently
diffusing and does not require the entire network to be taken into memory.

\begin{table}[ht!]
\centering
\small{
\begin{tabular}{|l|c|c|c|c|}
\hline
{\bf Algorithm} & {\bf Precision} & {\bf Recall} & {\bf F-meas.} & {\bf
AUC} \\
\hline
Random    &   10.00        & 50.0        & 16.66       & 10.0 \\
CGNP(E1)  &   26.07        & 59.59       & 36.28       & 28.7 \\
CGNP(E)   &   29.50        & 50.65       & 37.28       & 33.8 \\ \hline
CGNP(E+N) &   32.22        & 52.51       & 39.94       & 39.1 \\
CGNP(E+C) &   46.12        & 54.37 & 49.91       & 52.5 \\
CGNP(E+N+C) & 43.61        & {\bf 61.63}       & {\bf 51.08}       & {\bf
53.1} \\
WFVP       &  {\bf 66.00}  & 36.00        & 46.58 &  - \\
\hline
\end{tabular}
}
\caption{Results comparing CGNP using various feature combinations with
WFVP on WengDat (all values in \%)}
\label{table:results-wengdat}
\end{table}

\subsection{Geographical Trends}
We also evaluated the performance of CGNP over hashtags based on their
spread within individual geographies, i.e., the graph of interest was
restricted to the nodes lying within individual geographies in the
dataset. In particular, we experimented with three different
geographies, namely, London, India and Quito. Since the number of
users across each geography varies, we appropriately scaled the
prediction and virality thresholds for individual geographies.  Virality
threshold was maintained at 10 times the prediction threshold, in line
with the ratio used for the entire dataset. The prediction threshold
was hand tuned to ensure there was sufficient information in the data
up to that point.
Table~\ref{table:geo-stats} presents the details about number of
active users (i.e. those who have tweeted at least once), prediction
and virality thresholds, and \% of viral hashtags for each of the above
geographies. Table~\ref{table:geo-results} presents the F-measure and
AUC values for various feature combinations for each of the
geographies. Note that since we are already {\em within} individual
geographies, the feature category $G$ is absent in the combination. As
seen in case of the full dataset, the performance increases with
increasing sophistication in the feature set. For London and India, 
maximum benefit is obtained using the conductance based features.
Network based features help improve this further. For Quito, network
based features seem to give a larger gain. Best feature combination 
is still E+N+C.

\begin{table}
\small
\centering
\begin{tabular}{|l|r|r|r|r|}
\hline
{\bf Geogr-} & {\bf \# Active} & {\bf Prediction} & {\bf Virality} & {\bf \% of Viral} \\
{\bf aphy}   & {\bf users}  & {\bf Threshold} & {\bf Threshold} & {\bf Hashtags}       \\
\hline
London & 226906 & 150 & 1500  & 3.56\\
India & 28935   & 100 & 1000  & 8.67\\
Quito & 91871   & 50  &  500  & 3.14\\
\hline
\end{tabular}
\caption{Individual Geography Statistics}
\label{table:geo-stats}
\end{table}
\begin{table}[ht!]
\centering
\small{
\begin{tabular}{|l|rr|rr|rr|}
\hline
             & \multicolumn{2}{c|}{London} & \multicolumn{2}{c|}{India} & \multicolumn{2}{c|}{Quito} \\
\hline
{\bf Algorithm} & {\bf F} & {\bf AUC} & {\bf F} & {\bf AUC} & {\bf F} & {\bf AUC}  \\
\hline
Random      \hspace{-0.1in} &  6.7       \hspace{-0.1in}    &  3.6        & 14.8      \hspace{-0.1in} &  8.7         & 5.9       \hspace{-0.1in} &  3.1         \\ 
CGNP(E1)     \hspace{-0.1in} & 14.53       \hspace{-0.1in}    & 9.1        & 26.85     \hspace{-0.1in} & 19.6         & 19.42       \hspace{-0.1in} & 8.8         \\
CGNP(E)     \hspace{-0.1in} & 15.38      \hspace{-0.1in}    & 10.4 & 31.26      \hspace{-0.1in} & 23.9         &	17.69       \hspace{-0.1in} & 9.0         \\ \hline
CGNP(E+N)     \hspace{-0.1in} & 14.21       \hspace{-0.1in}    & 9.3        & 34.84      \hspace{-0.1in} & 27.6         & 22.52       \hspace{-0.1in} & 17.1         \\
CGNP(E+C)     \hspace{-0.1in} & 20.19       \hspace{-0.1in}    & 14.8  & 37.54      \hspace{-0.1in} & 31.3         & 17.42       \hspace{-0.1in} & 9.7         \\
CGNP(E+N+C) \hspace{-0.1in} & {\bf 22.17} \hspace{-0.1in}      & {\bf 15.5}  & {\bf 42.03}\hspace{-0.1in} & {\bf 36.2}   & {\bf 23.56} \hspace{-0.1in} & {\bf 17.3}   \\
\hline
\end{tabular}
}
\caption{Results on Individual Geographies.}
\label{table:geo-results}
\end{table}

\section{Conclusions}
\label{sec:conclusions}
In this work, we have carefully studied the effect of three different
sets of feature categories, i.e., network based, geography based and
conductance based, for the task of predicting hashtag virality in a
large dataset. Our main contribution is a novel feature set that
includes new features based on the network properties of the users
tweeting a hashtag and the geographical information contained in their
profiles and in their tweets. Building on the intuition that the
spread of memes across communities is a critical discriminator of
viral topics we have introduced a suite of conductance based features
for the prediction task.  We found that all our three feature
categories (apart from the baseline evolution based features), have a
significant impact on virality prediction, with conductance being the
most effective. This justifies the intuition regarding the
relationships of communities and virality and suggests that a more
dynamic view of communities, centred around the diffusion pattern of
individual hashtags, is more appropriate and effective for the
prediction task. The fact that our feature set outperforms approaches
relying on static communities detected in the network (such as the
work of~\cite{weng&al13a}) is doubly important in view of the fact
that detecting static communities in the entire network is
very expensive computationally at scale.

Future research directions include further investigating the use of
our proposed feature sets for predicting spread of topics in
individual geographies, more carefully examining the relative impact
of community based and conductance based features and incorporating
semantic features in our framework.

\bibliographystyle{plain} 
\bibliography{virality}

\begin{thebibliography}{10}

\bibitem{agarwal-iitd:2013}
Pranay Agarwal.
\newblock Prediction of trends in online social network.
\newblock Master's thesis, Indian Institute of Technology, Delhi, 2013.

\bibitem{Aral}
Sinan Aral and Dylan Walker.
\newblock Creating social contagion through viral product design: A randomized
  trial of peer influence in networks.
\newblock {\em Manag. Sci.}, 57(9):1623--1639, 2011.

\bibitem{Ardon}
Sebastien Ardon, Amitabha Bagchi, Anirban Mahanti, Amit Ruhela, Aaditeshwar
  Seth, Rudra~Mohan Tripathy, and Sipat Triukose.
\newblock Spatio-temporal and events based analysis of topic popularity in
  {Twitter}.
\newblock In {\em Proc.~22nd ACM Intl. Conf. on Information and Knowledge
  Management (CIKM 2013)}, pages 219--228. ACM, 2013.

\bibitem{berger}
Jonah Berger and Katherine~L Milkman.
\newblock What makes online content viral?
\newblock {\em J. Marketing Res.}, 49(2):192--205, 2012.

\bibitem{bright-arstechnica:2011}
Peter Bright.
\newblock How the london riots showed us two sides of social networking.
\newblock Posted on http://arstechnica.com/, 11 August 2011, August 2011.

\bibitem{cheng-www:2014}
Justin Cheng, Lada~A. Adamic, P.~Alex Dow, Jon~M. Kleinberg, and Jure Leskovec.
\newblock Can cascades be predicted?
\newblock In {\em Proc.~23rd Intl. World Wide Web Conference (WWW '14)}, pages
  925--936, 2014.

\bibitem{chierichetti-stoc:2010}
F.~Chierichetti, S.~Lattanzi, and A~Panconesi.
\newblock Almost tight bounds for rumour spreading with conductance.
\newblock In {\em Proc.~42nd ACM Symp. on Theory of computing (STOC '10)},
  pages 399--408, 2010.

\bibitem{davis2006relationship}
Jesse Davis and Mark Goadrich.
\newblock The relationship between precision-recall and roc curves.
\newblock In {\em Proceedings of the 23rd international conference on Machine
  learning}, pages 233--240. ACM, 2006.

\bibitem{drummond&holte03}
Chris Drummond and Robert~C. Holte.
\newblock C4.5, class imbalance, and cost sensitivity: Why under-sampling beats
  over-sampling.
\newblock In {\em ICML Workshop on Learning from Imbalanced Datasets}, pages
  1--8, 2003.

\bibitem{ghosh-wsdm:2011}
R.~Ghosh and K.~Lerman.
\newblock A framework for quantitative analysis of cascades on networks.
\newblock In {\em Proc.~4th ACM International Conference on Web search and data
  mining (WSDM '11)}, pages 665--674, 2011.

\bibitem{gleich-kdd:2012}
David~F. Gleich and C.~Seshadhri.
\newblock Vertex neighborhoods, low conductance cuts, and good seeds for local
  community methods.
\newblock In {\em Proc.~18th ACM SIGKDD Intl. Conf. on Knowledge Discovery and
  Data Mining (KDD '12)}, pages 597--605, 2012.

\bibitem{guerini-socialcom:2013}
Marco Guerini, Jacopo Staiano, and David Albanese.
\newblock Exploring image virality in {Google Plus}.
\newblock In {\em Proc.~ ASE/IEEE Intl. Conference on Social Computing
  (SocialCom 2013)}, pages 671--678, 2013.

\bibitem{guerini}
Marco Guerini, Carlo Strapparava, and G{\"o}zde {\"O}zbal.
\newblock Exploring text virality in social networks.
\newblock In {\em Proc.~Intl. AAAI Conf. on Weblogs and Social Media (ICWSM
  2011)}, 2011.

\bibitem{guruswami-unpub:2000}
Venkatesan Guruswami.
\newblock Rapidly mixing {Markov} chains: A comparison of techniques.
\newblock Available at: http://www.cs.cmu.edu/~venkatg/pubs/pubs.html, 2000.

\bibitem{Hansen}
Lars~Kai Hansen, Adam Arvidsson, Finn~{\AA}rup Nielsen, Elanor Colleoni, and
  Michael Etter.
\newblock Good friends, bad news-affect and virality in twitter.
\newblock In {\em Future information technology}, pages 34--43. Springer, 2011.

\bibitem{harenberg-wics:2014}
Steve Harenberg, Gonzalo Bello, L.~Gjeltema, Stephen Ranshous, Jitendra
  Harlalka, Ramona Seay, Kanchana Padmanabhan, and Nagiza Samatova.
\newblock Community detection in large-scale networks: a survey and empirical
  evaluation.
\newblock {\em WIREs Comput Stat}, 6:426--439, 2014.

\bibitem{jenders-msnd:2013}
Maximilian Jenders, Gjergji Kasneci, and Felix Naumann.
\newblock Analyzing and predicting viral tweets.
\newblock In {\em WWW (Companion Volume)}, pages 657--664, 2013.

\bibitem{jerrum-siamjc:1989}
M.~R. Jerrum and A.~J. Sinclair.
\newblock Approximating the permanent.
\newblock {\em SIAM J. Comput.}, 18:1149--1178, 1989.

\bibitem{Kitsak}
Maksim Kitsak, Lazaros~K Gallos, Shlomo Havlin, Fredrik Liljeros, Lev Muchnik,
  H~Eugene Stanley, and Hern{\'a}n~A Makse.
\newblock Identification of influential spreaders in complex networks.
\newblock {\em Nature Phys.}, 6(11):888--893, 2010.

\bibitem{kulshrestha-icwsm:2012}
Juhi Kulshrestha, Farshad Kooti, Ashkan Nikravesh, and Krishna~P. Gummadi.
\newblock {Geographic Dissection of the {Twitter} Network}.
\newblock In {\em Proc.~ICWSM 2012}, 2012.

\bibitem{Kwak_2010}
H.~Kwak, C.~Lee, H.~Park, and S.~Moon.
\newblock What is {Twitter}, a social network or a news media?
\newblock In {\em Proc.~19th Intl. conference on World Wide Web (WWW '10)},
  pages 591--600, 2010.

\bibitem{Lermann}
Kristina Lerman and Tad Hogg.
\newblock Using a model of social dynamics to predict popularity of news.
\newblock In {\em Proc.~19th Intl. conference on World Wide Web (WWW '10)},
  pages 621--630. ACM, 2010.

\bibitem{Leskovec_2009}
J.~Leskovec, L.~Backstrom, and J.~Kleinberg.
\newblock Meme-tracking and the dynamics of the news cycle.
\newblock In {\em Proc.~KDD '09}, pages 497--506. ACM, 2009.

\bibitem{leskovec-www:2010}
Jure Leskovec, Kevin~J. Lang, and Michael~W. Mahoney.
\newblock Empirical comparison of algorithms for network community detection.
\newblock In {\em Proc.~19th Intl. Conf. on World Wide Web (WWW '10)}, pages
  631--640, 2010.

\bibitem{ma-jasist:2013}
Zongyang Ma, Aixin Sun, and Gao Cong.
\newblock On predicting the popularity of newly emerging hashtags in twitter.
\newblock {\em J. Assoc. Inf. Sci. Technol.}, 64(7):1399--1410, 2013.

\bibitem{mcgee-cikm:2013}
J.~McGee, J~Caverlee, and Z.~Cheng.
\newblock Location prediction in social media based on tie strength.
\newblock In {\em Proc.~22nd ACM Intl. Conf. on Information and Knowledge
  Management (CIKM 2013)}, pages 459--468, 2013.

\bibitem{myers-kdd:2012}
Seth~A. Myers, Chenguang Zhu, and Jure Leskovec.
\newblock Information diffusion and external influence in networks.
\newblock In {\em Proc.~KDD '12}, pages 33--41, 2012.

\bibitem{oh-rumour:2013}
Onook Oh, Manish Agrawal, and H~Raghav Rao.
\newblock {\em Rumor and Communication in Asia in the Internet Age}, chapter~8,
  pages 143--155.
\newblock Taylor and Francis, 2013.

\bibitem{rajyalakshmi2012topic}
S~Rajyalakshmi, Amitabha Bagchi, Soham Das, and Rudra~M Tripathy.
\newblock Topic diffusion and emergence of virality in social networks.
\newblock {\em arXiv preprint arXiv:1202.2215}, 2012.

\bibitem{Romero_2011}
D.~M. Romero, B.~Meeder, and J.~Kleinberg.
\newblock Differences in the mechanics of information diffusion across topics:
  idioms, political hashtags, and complex contagion on twitter.
\newblock In {\em Proc.~20th Intl. conf. on World Wide Web (WWW '11)}, pages
  695--704, 2011.

\bibitem{Suh}
Bongwon Suh, Lichan Hong, Peter Pirolli, and Ed~H Chi.
\newblock Want to be retweeted? large scale analytics on factors impacting
  retweet in twitter network.
\newblock In {\em IEEE/ASE SocialCom 2010}, pages 177--184. IEEE, 2010.

\bibitem{Szabo}
Gabor Szabo and Bernardo~A Huberman.
\newblock Predicting the popularity of online content.
\newblock {\em Comm. ACM}, 53(8):80--88, 2010.

\bibitem{totti-websci:2014}
Luam~Catao Totti, Felipe~Almeida Costa, Sandra Eliza~Fontes de~Avila, Eduardo
  Valle, Wagner~Meira Jr., and Virgilio Almeida.
\newblock The impact of visual attributes on online image diffusion.
\newblock In {\em Proc.~ACM Web Science Conference (WebSci '14)}, pages 42--51,
  2014.

\bibitem{Weng2}
Lilian Weng, Alessandro Flammini, Alessandro Vespignani, and Filippo Menczer.
\newblock Competition among memes in a world with limited attention.
\newblock {\em Sci. Rep.}, 2, 2012.

\bibitem{weng&al13a}
Lilian Weng, Filippo Menczer, and Yong-Yeol Ahn.
\newblock Virality prediction and community structure in social networks.
\newblock {\em Sci. Rep.}, 3, 2013.

\bibitem{weng&al13}
Lilian Weng, Filippo Menczer, and Yong-Yeol Ahn.
\newblock Predicting successful memes using network and community structure.
\newblock In {\em 8th Intl. AAAI Conference on Weblogs and Social Media (ICWSM
  2014)}, 2014.

\bibitem{Fang}
Fang Wu and Bernardo~A Huberman.
\newblock Novelty and collective attention.
\newblock {\em Proc. Natl. Acad. Sci. U.S.A}, 104(45):17599--17601, 2007.

\bibitem{zaman-aas:2014}
T~Zaman, E~B Fox, and E~T Bradlow.
\newblock A bayesian approach for predicting the popularity of tweets.
\newblock {\em Ann. Appl. Stat.}, 8(3):1583--1611, 2014.

\end{thebibliography}

\end{document}